\documentclass[12pt]{article}
\usepackage{amsfonts,latexsym,fullpage}
\newtheorem{Definition}{Definition}
\newtheorem{Theorem}{Theorem}
\newtheorem{Lemma}{Lemma}
\newtheorem{Corollary}{Corollary}

\title{Simulating Arbitrary Pair-Interactions by a Given Hamiltonian:
Graph-Theoretical Bounds on the Time Complexity}
\author{P.~Wocjan, D.~Janzing, and Th.~Beth\\
Institut f{\"u}r Algorithmen und Kognitive Systeme,\\
Universit{\"a}t Karlsruhe\\
Am Fasanengarten 5, D--76\,131 Karlsruhe, Germany
}

\date{June 13, 2001}

\begin{document}

\maketitle

\abstract{
We use an $n$-spin system with permutation symmetric $zz$-interaction 
for simulating arbitrary pair-interaction  Hamiltonians. The calculation of 
the required time overhead is mathematically equivalent to a 
separability problem of $n$-qubit density matrices.
We derive lower and upper bounds in terms of chromatic index and the spectrum of 
the interaction graph. The complexity measure defined by such a computational model
is related to gate complexity and a continuous complexity measure introduced in a 
former paper. We use majorization of graph spectra for classifying Hamiltonians
with respect to their computational power.
}

\section{Introduction}
The most common models for quantum computers use single and two qubit gates
as basic transformations in order to generate arbitrary unitary operations
on the quantum registers. Most discussions about the generation of 
quantum algorithms, quantum codes and possible realizations had successfully been
based on this concept. Mostly, even the definition of quantum complexity refers 
to such a model \cite{nielsen}. Nevertheless there is a priori no reason, why 
two qubit gates should be considered as basic operations for future quantum 
computers. In principle every quantum system could serve as a quantum register 
provided that its time evolution can be controlled in a universal way. At first 
sight, every definition of quantum complexity seems hence to be adequate only 
for a specific model of quantum computation. But it seems to be  a rather 
general feature of Hamiltonians available in nature that particles interact with
other particles in such a form, that the total Hamiltonian is a sum of 
pair-interactions. Therefore we want to base quantum complexity 
theory only on such a general feature.\footnote{This might be seen in the spirit of 
D.~Deutsch's statement ``What computers can or cannot compute is determined by the 
laws of physics alone and not by pure mathematics.''\cite{nielsen}, Chapter II}
This feature justifies  the following control theoretic model: If $n$ qubits are 
assumed to be physically represented 
by $n$ particles, the only part of the system's Hamiltonian which can be changed 
by extern access is the free Hamiltonian of each qubit. These $1$-particle
Hamiltonians might be controllable since they are only effective Hamiltonians 
which are phenomenologically given by an interaction to many extern particles 
(mean-field approximation \cite{duffield}). Based on results of quantum control 
theory in $2$-spin systems \cite{kha} we investigate the problem of simulating 
arbitrary pair-interaction Hamiltonians by a given one. We assume that the 
Hamiltonian of the $n$-system is a permutation invariant $zz$-interaction and 
show that the computational power\footnote{in the sense of time required to 
generate unitaries} of this Hamiltonian (together with local 
transformations on each spin) is at least as large as the power of quantum 
computers with $2$ qubit gates. For infinitesimal time evolutions, it turns out 
to be even stronger. 

We develop a theory, where the  computational power of a Hamiltonian
for simulating arbitrary Hamiltonians is characterized by features of
the interactions graphs. Standard concepts of graph theory like chromatic index and
spectrum of the adjacency matrix together with majorization
turn out to provide lower and upper bounds on the simulation overhead. Here we
are interested in the exact overhead and not only in polynomial equivalence as
in \cite{dodd}.

\section{Our model of computation}
Based on the approach of \cite{kha} we consider the following model. The quantum
system is a spin system, i.e.\ its Hilbert space is 
$\mathcal{H}_n=(\mathbb{C}^2)^{\otimes n}$, and its
Hamiltonian $H_d\in\mathfrak{su}(2^n)$ consists only of pair-interactions, i.e.\
\begin{equation}
H_d=\sum_{1\le k<l\le n} H_{k,l}
\end{equation}
where $H_{kl}$ acts only on the Hilbert space of the qubits $k$ and $l$. We 
assume that for every $k$ and $l$ the Hamiltonian $H_{kl}$ describes a 
non-trivial coupling and is traceless. The system's Hamiltonian $H_d$ is also 
called the \emph{drift} Hamiltonian since it is always present. We assume that 
we can perform all unitaries in the \emph{control} group 
$K=SU(2)\otimes\cdots\otimes SU(2)$ arbitrarily fast compared to the 
time evolution of the internal couplings between the qubits. Let $G$ be the 
unitary Lie group $SU(2^n)$ and $u\in G$ be a unitary we want to realize.
To achieve this all we can do is perform $v_1\in K$, wait $t_1$, perform
$v_2\in K$, wait $t_2\,,\ldots,$ perform $v_p\in K$ and wait $t_p$. The 
resulting unitary is 
$$u=\exp(i H_d t_p) v_p \cdots \exp(i H_d t_2) v_2 \exp(i H_d t_1) v_1\,. $$
This can be written as
$$ u=k_p \exp(i k_p^\dagger H_d k_p) \cdots \exp(i k_2^\dagger H_d k_2) 
         \exp(i k_1^\dagger H_d k_1)$$ 
where $k_i=v_i \cdots v_1$ for $i=1,\ldots,p$. This is just the solution of a 
time-dependent Schr\"odinger equation with piecewise constant Hamiltonians 
-- conjugates of the drift Hamiltonian $H_d$ by unitaries of $K$ -- followed by 
the unitary $k_p\in K$.
Let $Ad_K(H_d)$ denote the conjugacy class 
$$ Ad_K(H_d)=\{Ad_k(H_d)=k^{\dagger} H_d k\mid k\in K\}\,. $$
\begin{Definition}
A continuous time algorithm $A$ of running time $T$ is a piecewise constant
function $t\mapsto H(t)$ from the interval $[0,T]$ onto the set 
$Ad_K(H_d)$ followed by some local unitary $k\in K$. We say $A$ implements $u$ 
if $u=k u(T)$ where $(u(t))_{t\in [0,T]}$ is the solution of the 
time-dependent Schr\"odinger equation 
$(d/dt) u(t) = -iH(t) u(t)$ with $u(0)=I$.
\end{Definition}
The complexity of a unitary in this model is the running time of the optimal 
continuous time algorithms. 

Let $\sigma_\alpha^i$ denote the Pauli spin matrix $\sigma_\alpha$ 
($\alpha=x,y,z$) that acts on the $i$th spin. For simplicity, we 
assume that the drift Hamiltonian is 
\begin{equation}
H_d=\sum_{1\le k<l\le n} \sigma_z^k\sigma_z^l\,.
\end{equation}
The physical systems we have in mind might be for example solid states
with long-range interactions. Of course one might object that the interaction 
strength always decreases with the distance between the interacting particles. It 
will turn out that the assumption on non-decreasing interaction strengths makes our
model rather strong with respect to its computational power. One should 
understand our assumptions as the attempt to use a strong computational model
which is still physically justificable. Many aspects of our theory can be 
developed in strong analogy for more general drift Hamiltonians.

In Section~\ref{LowUpBounds} we will compare the computational power 
of our model with the power of quantum computers based on $2$-qubit gates.
Our arguments refer always to infinitesimal time evolutions, i.e.,
we will show that our model can implement quantum gates without overhead since we 
can simulate the time evolution implementing parallelized quantum gates.

First we have to define what we mean by
simulating  the time evolution $\exp(i H t)$ 
during a small time interval $[0,\epsilon]$ where $H$ is an arbitrary
pair-interaction Hamiltonian. Assume we have written $H$ as a positive linear 
combination $H=\sum_j \mu_j H_j$ with $\mu_j>0$ and each $H_j$ is an element 
of the conjugacy class $Ad_K(H_d)$. For small $\epsilon$ the unitary 
\[
\prod_j \exp(i\epsilon\mu_j H_j)
\]
is a good approximation for 
\[
\exp(i\epsilon H)=\exp(i\epsilon\sum_j\mu_j H_j)\,.
\]
This approximation is implemented if the system evolves the time 
$\epsilon\mu_j$ with respect to the Hamiltonian $H_j$.
The sum $\mu=\sum_j\mu_j$ is exactly the time overhead of the simulation.
Hence the problem is to express $H$ as a positive linear combination 
such that the overhead $\mu$ is minimal. Of course such a procedure might not be
optimal if one were interested in the implementation of $\exp(iHs)$ for any 
special value of $s$. Here we want to imitate the whole dynamical time evolution
$(\exp(iHs))_{s>0}$ in arbitrary small steps $\epsilon$. Then the optimization
reduces clearly to the convex problem stated above.

In the following it will be convenient to use a concise representation for 
the drift Hamiltonian and the interaction to be simulated:
a pair interaction Hamiltonian between qubits $k$ and $l$ can be written as
\begin{equation}
H_{kl} = \sum_{\alpha,\beta=x,y,z} J_{kl;\alpha\beta} 
\sigma_{\alpha}^k\sigma_{\beta}^l\,.
\end{equation}
The strengths of the components are represented by the pair-interaction
matrix
\begin{equation}
J_{kl}=\left(
\begin{array}{ccc}
J_{kl;xx} & J_{kl;xy} & J_{kl;xz} \\
J_{kl;yx} & J_{kl;yy} & J_{kl;yz} \\
J_{kl;zx} & J_{kl;zy} & J_{kl;zz}
\end{array}
\right)\in\mathbb{R}^{3\times 3}\,.
\end{equation}
The total Hamiltonian $H$ is represented by the $J$-matrix
\begin{equation}
J=\left(
\begin{array}{c|c|c|c|c}
0      & J_{12} & J_{13} & \cdots & J_{1n} \\ \hline
J_{21} & 0      & J_{23} & \cdots & J_{2n} \\ \hline
J_{31} & J_{32} & 0      &        & J_{3n} \\ \hline
\vdots & \vdots &        & \ddots &        \\ \hline
J_{n1} & J_{n2} & J_{n3} & \quad & 0
\end{array}
\right)\in\mathbb{R}^{3n\times 3n}\,.
\end{equation}

To explain more explicitly, why our simulation  problem is
a convex optimization, we recall that
every convex combination $\mu H_1 + (1-\mu)H_2$
of two Hamiltonians $H_1$ and $H_2$  can be 
simulated with overhead $1$ if $H_1$ and $H_2$ can.
Remarkably, the  problem of specifying
the set of Hamiltonians which can be simulated with overhead $1$
is  related to the problem of 
generalizing Bell inequalities to $n$-qubit states.  
More specifically, the convex problem can be reduced to the question
`how strong can $2$-spin correlations be in a separable $n$-qubit quantum
state?'

\begin{Theorem}[Optimal simulation]\label{optimal}
The Hamiltonian $H$ can be simulated with overhead $\mu$ if and only if there
is a separable quantum state $\rho$ in $(\mathbb{C}^2)^{\otimes n}$ such that
$$ \frac{1}{\mu}J+I=
(\mathrm{tr}(\rho\sigma_{\alpha}^k\sigma_{\beta}^l))_{kl;\alpha\beta}$$
where $J$ denotes the $J$-matrix of $H$ and $I$ the $3n\times 3n$ identity matrix.
\end{Theorem}

Proof:
By rescaling the considered Hamiltonian, it is sufficient to show that this is 
true for all Hamiltonians in $Ad_K(H_d)$ with $\mu=1$.
Assume we have written $H$ as a convex combination $H=\sum_j\mu_j H_j$
with $H_j\in Ad_K(H_d)$. In order to show that there is a separable state
of the desired form it is sufficient to show that the $J$-matrix of each 
Hamiltonian $H_j$ satisfies the equation of the theorem for an appropriate 
separable state. Let $H_j=uH_du^\dagger$. $H_j$ can be represented by 
$n$ three dimensional real unit vectors: to each qubit we associate the vector
$|J_k\rangle=(J_{k;x},J_{k;y},J_{k;z})^t\in \mathbb{R}^3$ where 
$u_k\sigma_z u_k^\dagger=J_{k;x}\sigma_x + J_{k;y}\sigma_y + J_{k;z}\sigma_z$ and
$u=u_1\otimes\ldots\otimes u_n$. The pair-interaction matrices are given 
by the matrix products $J_{kl}=|J_k\rangle\langle J_l|$.

By the Bloch sphere representation we have a correspondence between the
unit vectors $|J_k\rangle$ and the projections 
$\rho_k$ in $\mathbb{C}^2$ defined by 
$J_{k;\alpha}=\mathrm{tr}(\rho_k \sigma_{\alpha})$. Let
$\rho$ be the product state
$\rho:=\rho_1\otimes\ldots\otimes\rho_n$. Then we have 
$J_{kl;\alpha\beta}=\mathrm{tr}(\rho \sigma_{\alpha}^k\sigma_{\beta}^l)$ 
for all
$k\neq l$. 
Note that the product of two different Pauli matrices is the third 
Pauli matrix multiplied by a scalar. The only problem that remains is
that we may have 
$\mathrm{tr}(\rho\sigma_\alpha^k\sigma_\beta^k)\neq 0$ for $\alpha\neq\beta$. 
We substitute $\rho$ by a state $\bar{\rho}$ in such a way that the 
expectation values of all traceless $1$-qubit observables vanish and the 
expectation values of all considered $2$-qubit observables remain unchanged. For
every $|J_k\rangle$ we can find $U'_k\in SO(3)$ such that 
$U'_k|J_k\rangle=-|J_k\rangle$. This rotation 
corresponds to conjugation of the qubit $k$ by a unitary $u'_k$. To
$-|J_k\rangle$ corresponds the projection
$\rho'_k := I_2-\rho_k$. Let
$$ \bar{\rho}:= \frac{1}{2} 
   (\rho_1\otimes\ldots\otimes\rho_n + \rho'_1\otimes\ldots\otimes\rho'_n)
\,. $$
Then we have 
$\mathrm{tr}(\bar{\rho} \sigma_{\alpha}^k\sigma_{\beta}^l)=
\mathrm{tr}(\rho \sigma_{\alpha}^k\sigma_{\beta}^l)$ for
all $k\neq l$ (all vectors are multiplied by $-1$ and therefore there is no 
effect on the pairs) and $J_{kk}$ is the $3\times 3$ identity matrix.

Assume conversely we have a separable state of the desired form. Take its 
decomposition into pure product states. By the Bloch sphere representation we 
obtain the required conjugations of the drift Hamiltonian $H_d$. The time they 
have to be applied are given by the coefficients in the convex decomposition.
\hfill $\Box$

\section{Lower and upper bounds}\label{LowUpBounds}
A simple lower bound on the simulation time overhead can be derived 
from the fact that $J/\mu+I$ has to be a positive matrix, which is
an easy conclusion from Theorem~\ref{optimal}. 

\begin{Corollary}[Lower bound] 
The absolute value of the smallest eigenvalue of the $J$-matrix is a 
lower bound on the simulation overhead of $H$.
\end{Corollary}

Proof: The matrix 
$$ (\mathrm{tr}(\rho\sigma_{\alpha}^k\sigma_{\beta}^l))_{kl;\alpha\beta} $$
is positive for every  state $\rho$ in $(\mathbb{C}^2)^{\otimes n}$:
let $|d\rangle=(d_{k;\alpha})$ be an arbitrary vector and
$A=\sum_{k,\alpha} d_{k;\alpha} \sigma_{\alpha}^k$. Then we have
$$ \sum_{k,l,\alpha,\beta} d_{k;\alpha} 
\mathrm{tr}(\rho\sigma_{\alpha}^k\sigma_{\beta}^l) d_{l;\beta} =
\mathrm{tr}(\rho AA^*)\ge 0\,. $$
{} \hfill $\Box$

Now we show that our computational model is at least as powerful as the usual
model with $2$-qubit gates, even if also cares about constant overhead.
We describe here briefly the quantum circuit model and introduce the
\emph{weighted depth} following \cite{jan}. It is a complexity measure for unitary 
transformations based on the quantum circuit model. We assume that two qubit gates 
acting on disjoint pairs of qubits can be implemented simultaneously and define:

\begin{Definition}
A {\em quantum circuit} $A$ {\em of depth} $k$ is a sequence of
$s$ steps $\{A_1,\dots,A_s\}$ where every step consists of a set
of two qubit gates  $\{u_{kl}\}_{k,l}$ acting on disjoint
pairs $(k,l)$ of qubits.
Every step $i$ defines a unitary operator $v_i$ by taking the product
of all corresponding unitaries in any order.
The product $u:=\Pi_{i\leq s} v_i$ is the `unitary operator implemented
by $A$'. 
\end{Definition}
The following quantity measures the deviation of a unitary operator
from the identity:
\begin{Definition}
The {\em angle} of an arbitrary unitary operator $u\in SU(4)$
is the smallest possible norm\footnote{Here $\|.\|$ denotes the operator norm 
given by $\|a\|:=\max_x \|ax\|$ where $x$ runs over the unit vectors of the 
corresponding Hilbert space.} $\|a\|$ of a self-adjoint operator 
$a\in\mathfrak{su}(4)$ which satisfies $\exp(ia)=u$.
\end{Definition}
It coincides with the time required for the implementation of $u$ if
the norm of the used Hamiltonian is $1$. We consider only the angle of 
two-qubit gates, i.e.\ we do not include the angle of local gates in the 
definition of the weighted depth.
The notion of angle allows us to formulate a modification of the term `depth'
which will later turn out to be decisive in connecting complexity measures
of discrete and continuous algorithms:
\begin{Definition}
Let $\alpha_i$ be the maximal angle of the unitaries performed
in step $i$. Then the {\em weighted depth} is defined to be
the sum $\alpha=\sum_i\alpha_i$.
\end{Definition}

Assuming that the implementation time of a unitary is proportional to its 
angle, the weighted depth is the running time of the algorithm.
We first need two technical lemmas to show that 
such an algorithm can be simulated by our computational model with complete
$zz$-Hamiltonian without any time overhead.
\begin{Lemma}
Let $M$ be a set of qubit pairs, such that no two pairs
contain a common qubit. Then we can simulate
\begin{equation}
H_M=\sum_{(k,l)\in M} \sigma_z^k\sigma_z^l
\end{equation}
with overhead $1$.
\end{Lemma}

Proof: This has been noted in \cite{leung}. Theorem \ref{independent} proves
a more general statement. \hfill $\Box$
\begin{Lemma}
Let $H_d=\sigma_z\otimes\sigma_z$ be the drift Hamiltonian of a $2$-spin system.
All Hamiltonians $H\in\mathfrak{su}(4)$ can be simulated with overhead less than 
$\|H\|$.
\end{Lemma}

Proof: We first assume that $H$ contains no local terms, i.e.\
$H=\sum_{\alpha,\beta} J_{\alpha\beta} \sigma_{\alpha}\otimes\sigma_{\beta}$.
Let $J_{12}$ be the matrix representing $H$. Conjugation of $H$ by 
$k=u\otimes v\in SU(2)\otimes SU(2)$ corresponds to multiplication of $J_{12}$ by
$U\in SO(3)$ from the left and by $V\in SO(3)$ from the right. By the singular
value decomposition \cite{horn} there are $U,V\in SO(3)$ such that
$J_{12}=U\mbox{diag}(s_x,s_y,s_z)V$ where $s_x,s_y,s_z$ are the singular values of 
$J_{12}$. Equivalently, there is $k\in SU(2)\otimes SU(2)$ such that 
$kHk^\dagger=H_{s_x,s_y,s_z}$ where 
$H_{s_x,s_y,s_z}=
s_x\sigma_x\otimes\sigma_x+s_y\sigma_y\otimes\sigma_y+s_z\sigma_z\otimes\sigma_z$.
By computing the eigenvalues we see that 
$\|H_{s_x,s_y,s_z}\|=\sum_{\alpha} |s_\alpha|$.
The simulation time overhead can not be more than 
the right hand side since each term $s_\alpha \sigma_\alpha \otimes 
\sigma_\alpha$ can be simulated with overhead $s_\alpha$.

Let $H$ contain local terms, i.e.\
$H=\sum_{\alpha} J_{\alpha\alpha} \sigma_{\alpha}\otimes\sigma_{\alpha} +
1\otimes a + b\otimes 1$. We can split $H=H'+H''$ where $H'$ is the non-local part
and $H''$ the local one. By the Trotter formula we can simulate the parts 
independently. The simulation of $H''$ takes no time by assumption.
It remains to show that $\|H'\|\le\|H\|$. We may
assume that $H$ is invariant with respect to qubit permutation since 
$\|\frac{1}{2}H+\frac{1}{2}H_{ex}\|\le\|H\|$ where $H_{ex}$ is the Hamiltonian 
obtained from $H$ by exchanging the qubits. 
By conjugation we can obtain a 
Hamiltonian of the form 
$H=H_{s_x,s_y,s_z} + s (1\otimes\sigma_x + \sigma_x\otimes 1)$. 
By computing the
eigenvalues we see that $\|H_{s_x,s_y,s_z}\|\le\|H\|$. 
\hfill $\Box$

\begin{Corollary}\label{step}
Let $A=\{u_{kl}\}_{k,l}$ be a step of a quantum circuit and 
$\alpha$ its weighted depth. Then the $zz$-model can simulate the 
unitary implemented by $A$ with overhead $\alpha$.
\end{Corollary}

Proof: Let $M=\{(k,l)\}$ be the set of the pairs which the two-qubit gates act on. 
No two pairs in $M$ contain a common vertex and therefore we can simulate the
Hamiltonian $H_M=\sum_{(k,l)\in M}\sigma_z^k\sigma_z^l$ with overhead $1$.
Let $H_{kl}$ be the Hamiltonian of minimal norm such that $u_{kl}=\exp(i H_{kl})$
for every $(k,l)\in M$. Now we can simulate every $H_{kl}$ parallely with overhead 
less than $\|H_{kl}\|$ by conjugating $H_M$.
\hfill $\Box$

Our goal is to compare interactions with respect to the simulation 
complexity in our model given by the complete $zz$-interaction and the quantum 
circuit model. For doing so, we need some basic concepts of graph theory \cite{bol}.
A graph is an ordered pair 
$G=(V,E)$ with $V\subseteq\{1,2,\ldots, n\}$ and 
$E=\{e_1,e_2,\ldots,e_m\}\subseteq V\times V$.
Elements of $V$ are called vertices. They label the qubits. Elements of 
$E$ are called edges. They label the pair-interactions between the qubits. 
An edge $e=(k,l)$ is an ordered pair of vertices $k$ and $l$ called the ends 
of $e$. We consider only undirected graphs with no loops. To have a unique 
representation we require that $k<l$. Two distinct edges are called adjacent 
if and only if they have a common end vertex. A subset $M$ of the edge set $E$
is called independent if no two edges of $M$ are adjacent in $G$. A graph $G$ 
is called complete if every pair of distinct vertices of $G$ are adjacent in 
$G$; such a graph is denoted by $K_n$. Rephrased in this language, 
our drift Hamiltonian is of the form
$$ H_d = \sum_{(k,l)\in E(K_n)} \sigma_z^k\sigma_z^l$$
and is called in the following the complete $zz$-Hamiltonian.

\begin{Definition}
Let $H$ be an arbitrary pair-interaction Hamiltonian. For every non-negative real 
number $r$ we define the {\em interaction graph} $G_r$ as follows: 
Let the qubits $\{1,\dots,n\}$ label the vertices and let the edges
be all the pairs $(k,l)$ with the property $\|H_{k,l}\|> r$.
\end{Definition}
The chromatic index $\chi'$ is the minimum number of colors permitting an 
edge-coloring such that no two adjacent edges receive the same color or 
equivalently a partition $E=M_1\cup M_2\cup \ldots\cup M_{\chi'}$ into 
independent subsets of $E$. The following quantity turns out to be an upper bound 
on the overhead.
\begin{Definition}
We define the {\em weighted chromatic index} of $H$
\begin{equation}
\chi':=\int_0^\infty \chi'_r dr
\end{equation}
where $\chi'_r$ denotes the chromatic index of  $G_r$.
\end{Definition}

In a former paper \cite{jan} we have introduced
the weighted chromatic index as a complexity measure of the interaction.
This point of view has been justified by two arguments, where
the first one is an observation in \cite{jan}:

\begin{Theorem}\label{janInfinite}
The evolution generated by a pair-interaction Hamiltonian $H$
during the infinitesimal time period $dt$ can be simulated by a parallelized 
$2$-qubit gate network with weighted depth $\chi'\, dt$ if $\chi'$ is the 
weighted chromatic index of $H$.
\end{Theorem}

The second argument to consider chromatic index as a complexity measure for the 
interaction is only intuitive: in general, it should be easy to control 
interactions on disjoint qubit pairs, whereas one should expect that its unlikely
that one can {\it control} simultaneously the interaction between qubit $1$ and $2$
and the interaction $1$ and $3$ at the same moment.
This `a priori'-assumption of \cite{jan} 
can be partly justified by the following corollary which is 
an easy conclusion of Corollary~\ref{step} and Theorem~\ref{janInfinite}.

\begin{Corollary}
The time overhead for simulating the Hamiltonian $H$ in the $zz$-model is at most 
the weighted chromatic index of $H$.
\end{Corollary}

The assumption that the drift Hamiltonian contains only pair-interaction of the
form $\sigma_z\otimes\sigma_z$ can be dropped. Let 
$H=\sum_{\alpha,\beta} J_{\alpha\beta} \sigma_\alpha\otimes\sigma_\beta$ be an 
arbitrary pair-interaction. By conjugating $H$ with 
$\{I\otimes I,I\otimes\sigma_z,\sigma_z\otimes I,\sigma_z\otimes\sigma_z\}$ we
obtain $J_{zz} \sigma_z\otimes\sigma_z$. This can be done with overhead $1$. The 
bounds of the corollary must be divided by the minimum $J_{zz}$ of all 
pair-interactions occurring in $H$.

\section{Applications}
The graph theoretical nature of our optimization problems
becomes even stronger if we reduce our attention to one type 
of interactions, namely $zz$-interactions. Then the desired Hamiltonian 
is completely described by a weighted graph.

We consider the problem to simulate the time evolution
\begin{equation}
H=\sum_{(k,l)} J_{kl} \sigma^k_z \sigma^l_z\,.
\end{equation}
when the complete $zz$-Hamiltonian is present. We first show that in this case 
it is sufficient to use conjugation by $\sigma_x$ only. Let 
$H':=\sigma_z\otimes\sigma_z$. Note that 
$(\sigma_x\otimes I) H' (\sigma_x\otimes I)=-H'$ and
$(\sigma_x\otimes\sigma_x) H' (\sigma_x\otimes\sigma_x)=H'$. In the following 
we denote conjugation by $\sigma_x$ by $-$ and no conjugation by $+$.
The Hamiltonian to be simulated contains only terms of the form 
$J_{kl;zz} \sigma^k_z \sigma_z^l$ by assumption. If it is written as a convex
combination of elements of $Ad_K(H_d)$ it is sufficient to show that
for each of these elements
there is a procedure which cancels the terms $J_{kl;\alpha\beta}$
for $(\alpha, \beta)\neq (z,z)$ without any effect on the $J_{kl,zz}$ terms.
Therefore consider $\tilde{H}=kH_dk^\dagger$. Then we can also achieve the 
Hamiltonian 
$\tilde{H}_{zz}=\sum_{(k,l)} \tilde{J}_{kl;zz}\sigma^k_z\sigma^l_z$ with 
overhead $1$.
For every qubit $i$ there is a $\tilde{J}_{i;z}$ such that 
$\tilde{J}_{kl;zz}=\tilde{J}_{k;z} \tilde{J}_{l;z}$ for
all edges $(k,l)$. We express each $\tilde{J}_{i;z}=c^{+}_i-c^{-}_i$ with 
$0\le c^+_k,c^-_k\le 1$ and $c^+_k+c^-_k=1$. Let 
$K=\{I,\sigma_x\}\otimes\ldots\otimes\{I,\sigma_x\}$. We conjugate
the drift Hamiltonian by $u=u_1\otimes u_2\otimes\ldots\otimes u_n\in K$ for
time $t(u)=\prod_{i=1}^n c_i(u)$ where $c_i(u)=c^+_k$ if $u_i=I$ and
$c_i(u)=c^-_k$ if $u_k=\sigma_x$. We have
$$ \sum_{u\in K} t(u) u\tilde{H}u^{\dagger} = \tilde{H}_{zz}\,. $$ 

Since we restrict our attention to interactions with $zz$-terms only
a shorter notation will be useful. 
To each edge $e=(k,l)$ of $G$, we associate a real number $w_{kl}$ called the 
weight of $e$. The resulting graph is called a weighted graph. Its adjacency 
matrix $J$ is the real symmetric matrix with zeros on the diagonal defined by
\begin{equation}
J_{ii}:=0\,,\quad J_{kl}:=w_{kl} \mbox{ and } J_{lk}:=w_{kl} 
\end{equation}
for all edges $(k,l)$ of $G$. 
An unweighted graph can be considered as a weighted whose edges all have the
weight $1$. 

A (unweighted) graph is bipartite if its vertex set can be partitioned into 
two nonempty subsets $X$ and $Y$ such that each edge of $G$ has one end in $X$
and the other in $Y$. The pair $(X,Y)$ is called a bipartition of the 
bipartite graph. The complete 
bipartite graph with bipartition $(X,Y)$ is denoted by 
$G(X,Y)$. 

A Seidel matrix defines a modified adjacency matrix $S=(s_{kl})$ 
for (unweighted) graphs in the following way \cite{cvet}:
$$
s_{kl}=\left\{
\begin{array}{rl}
-1 & \mbox{ if $k$ and $l$ are adjacent } k\neq l \\
 1 & \mbox{ if $k$ and $l$ are non-adjacent} \\
\end{array}
\right.
$$
and $s_{kk}=0$. Obviously, $S=K-I-2J$, where $K$ denotes a square matrix all
of whose entries are equal to $1$ and $J$ the adjacency matrix of $G$.

\begin{Theorem}[Optimal simulation]
A graph $G$ can be simulated with overhead $1$ if and only if it can be 
expressed as a convex combination
\begin{equation}
J=\sum_i t_i S_i
\end{equation}
where the sum  runs over the Seidel adjacency matrices of all complete
bipartite graphs, i.e., over $2^{n-1}$ possible matrices.
\end{Theorem}

Proof: By assigning to each vertex either $+$ or $-$ we have a bipartition
of the vertex set: $X$ contains all vertices with $+$ and $Y$ all vertices with 
$-$. The sign of the edge $(k,l)$ is $-$ if and only if the edge has one end 
in $X$ and the other end in $Y$ and $+$ otherwise. The edges with $-$ define 
the complete bipartite graph $G(X,Y)$. We also include the case $X=\emptyset$ 
and $Y=V$ to cover the case when $+$ is assigned to all knots. 
Therefore all we can achieve in the one step is $K-I-2J(X,Y)$.
\hfill $\Box$

\begin{Corollary}[Lower bound]
The absolute value of the smallest eigenvalue of the $J$-matrix is a lower 
bound on the simulation overhead.
\end{Corollary}

We present now some upper bounds on the overhead.
A graph $G=(V',E')$ is called a subgraph of $G'$ if $V'\subseteq V$ and
$E'\subseteq E$. A clique of $G$ is a complete subgraph of $G$. A clique of 
$G$ is called a maximal clique of $G$ if it is not properly contained in 
another clique of $G$.  A clique partition $P$ of $G$ is a partition of
$E(G)$ such that its classes induce maximal cliques of $G$. Given a set 
$C$ of $h$ colors, an $h$-coloring of $P$ in $G$ is a mapping from $P$ to $C$,
such that cliques sharing a vertex have different colors. Let the clique coloring
index $c(G)$ be the smallest $h$ such that there is a partition $P$ permitting 
an $h$-coloring \cite{wallis}. We say the graph $G$ consists of independent 
cliques if $c(G)=1$.  
\begin{Lemma}[Upper bound]\label{independent}
Let $G$ be a graph consisting of independent cliques.
We can simulate the Hamiltonian $H_E$ with overhead $1$ which is optimal.
\end{Lemma}

Proof: Let $\omega\ge 2$ be the number of maximal cliques. We construct $\omega$ 
vectors $s_i$ of length $2^{\omega-1}$ as follows:
$$
\begin{array}{lcl}
s_1 & = & (++++++++\cdots) \\
s_2 & = & (+-+-+-+-\cdots) \\
s_3 & = & (++--++--\cdots) \\
    & \vdots & \\
s_\omega & = & 
(\underbrace{++\,\cdots\,+}_{2^{\omega-2}}
 \underbrace{--\,\cdots\,-}_{2^{\omega-2}})\\
\end{array}
$$
where $+$ stands for $1$ and $-$ for $-1$. The scalar products are
$\langle s_i,s_j\rangle=2^{\omega-1} \delta_{ij}$. We partition the time 
interval into $2^{\omega-1}$ intervals of equal length. In the $m$th interval 
we conjugate all qubits of the $i$th clique by $\sigma_x$ if $s_{i,m}=-$ and
do nothing otherwise. This is optimal since $q\le -1$ where $q$ is the
smallest eigenvalue of $G$.\hfill $\Box$

The scheme used in the proof is time optimal. But the number of conjugations
grows exponentially with the number of cliques $\omega$. It is possible to use the 
conjugations schemes based on Hadamard matrices \cite{leung}. There the number 
of conjugations grows only quadratically with $\omega$.

\begin{Corollary}[Upper bound]
Let $G$ be an arbitrary graph. Then the Hamiltonian $H_G$ can be simulated 
with the overhead $c(G)$.
\end{Corollary}
Note that if $M$ is an independent set than the graph $G=(V,M)$ consists of 
independent cliques. Therefore the chromatic index
is an upper bound on clique index. However, this bounds is not always good. 
Consider e.g.\ the graph $G$ containing all edges that do not have $1$ as end 
vertex. Then the chromatic index of $G$ is still high but the clique coloring 
index is only $1$. 

Since the optimal simulation of graph consisting of independent cliques has 
overhead $1$ one might think that the clique index is the smallest overhead. 
But this is not so as shows the following example. Consider the star 
$G=(V,E)$ with $V=\{1,\ldots,5\}$ and $E=\{(1,2),(1,3),(1,4),(1,5)\}$. 
The clique index of is $4$ but the optimal simulation has overhead $2$ only. 
The vectors can be chosen as 
$s_1=(++++),\, s_2=(-+++),\, s_3=(+-++),\, s_4=(++-+),\, s_5=(+++-)$ and each 
of the four intervals has length $1/2$. This is optimal since the smallest 
eigenvalue of the adjacency matrix of $G$ is $-2$.

\section{Quasi-order of Hamiltonians}
Let $H$ and $\tilde{H}$ be arbitrary pair-interaction Hamiltonians. We 
investigate the question whether $\tilde{H}$ can be simulated by $H$ with overhead 
$\mu$. Note that this defines a quasi-order of the pair-interaction Hamiltonians 
for $\mu=1$. A partial characterization of the quasi-order is expressed
in terms of majorization of the spectra of the corresponding 
matrices $J$ and $\tilde{J}$. Similar methods have been used to derive 
conditions for a class of entanglement transformations 
and to characterize mixing and measurement in quantum mechanics 
\cite{nielsen1,nielsen2}.

Suppose that $x=(x_1,\ldots,x_d)$ and $y=(y_1,\ldots,y_d)$ are two dimensional
real vectors. We introduce the notation $\downarrow$ to denote
the components of a vector rearranged into non-increasing order, so
$x^\downarrow=(x_1^\downarrow,\ldots,x_d^\downarrow)$, where
$(x_1^\downarrow\ge x_2^\downarrow\ge\ldots\ge x_d^\downarrow)$. We say that
$x$ is majorized by $y$ and write $x\prec y$, if
$$ \sum_{j=1}^k x_j^\downarrow\le \sum_{j=1}^k y_j^\downarrow\,, $$
for $k=1,\ldots,d-1$, and with equality when $k=d$ \cite{bha}.

Let $\mathrm{Spec}(X)$ denote the spectrum of the hermitian matrix $X$, i.e.\ the 
vector of eigenvalues, and $\lambda(X)$ denote the 
vector of components of $\mathrm{Spec}(X)$ arranged so they appear in 
non-increasing order. Ky Fan's maximum principle \cite{nielsen2} states that for 
any Hermitian matrix $A$, the sum of the $k$ largest eigenvalues of $A$ is the 
maximum value $\mathrm{tr}(AP)$, where the maximum is taken over all 
$k$-dimensional projections $P$,
$$ \sum_{j=1}^k \lambda_j(A)=\max_P\mathrm{tr}(AP)\,. $$
It gives rise to a useful constraint on the eigenvalues of a sum of two 
Hermitian matrices $C:=A+B$, that $\lambda(C)\prec\lambda(A)+\lambda(B)$. Choose
a $k$-dimensional projection $P$ such that
\begin{equation}
\sum_{j=1}^k\lambda_j(C) = \mathrm{tr}(CP) = 
\mathrm{tr}(AP)+\mathrm{tr}(BP) \le 
\sum_{j=1}^k \lambda_j(A)+\sum_{j=1}^k \lambda_j(B)\,. \label{EigIneq}
\end{equation}
This permits us to derive a lower bound on the simulation overhead.
\begin{Lemma}[Majorization]
Let $H$ and $\tilde{H}$ be arbitrary pair-interaction Hamiltonians. A necessary
condition that $\tilde{H}$ can be simulated with overhead $\mu$ by $H$ is
that $\mathrm{Spec}(\tilde{J})\prec\mu\mathrm{Spec}(J)$. 
\end{Lemma}

Proof: By representing the Hamiltonians by their $J$-matrices we see that 
$\tilde{H}$ can be simulated with overhead $\mu$ if and only if there is
a sequence of orthogonal matrices 
$U_j=U_{j1}\oplus\ldots U_{jn}\in SO(3)\oplus\ldots\oplus SO(3)$ and 
$\mu_j>0$ with $\sum_j\mu_j=\mu$ such that
$$ \tilde{J}=\sum_j\mu_j U_j J U_j^T\,. $$
The proof now follows from the inequality~(\ref{EigIneq}).

We consider now the problem to reverse the time evolution $\exp(iH_d t)$, i.e.\
what is the overhead of simulating $-H_d$ when $H_d$ is present. 
\begin{Lemma}[Lower bound on inverting]
Let $r$ be the greatest eigenvalue and $q$ the smallest eigenvalue of $J$. 
Then $\mu\ge\frac{r}{-q}$ is a lower bound on the overhead for simulating $-H_d$ by 
$H_d$.
\end{Lemma}
Proof: This is a direct consequence of the Weyl inequality (see 
\cite{bha}, Theorem~III.2)
$\lambda_d(A+B)\ge\lambda_d(A)+\lambda(B)$ for the sum of two
Hermitian matrices where $\lambda_d$ denotes the smallest eigenvalue.
\hfill $\Box$

Let $G$ be a connected graph and 
$H_d=\sum_{(k,l)\in E(G)}\sigma_z^k\sigma_z^l$. If $G$ is not connected then
the components can be treated independently. For the spectrum the following
statements hold (see \cite{cvet}, Theorem~0.13):
$$ 1\le r\le n-1\,,\quad -r\le q\le -1\,. $$
It is interesting to note that this gives a tight bound for simulating $-H_d$ when
$G=K_n$ since for a complete graph we have $r=(n-1)$ and $q=-1$. An upper bound 
is the (weighted) chromatic index $\chi'(K_n)$ which is either $n$ 
(if $n$ is even) or $n-1$ (otherwise). This simple example shows that the inverse 
of the natural time evolution may have a relatively high complexity.

\begin{Lemma}
Let the drift Hamiltonian $H_d$ be an arbitrary pair-interaction Hamiltonian. If 
the interaction graph $G_0(H_d)$ is bipartite then we can invert the time 
evolution with overhead of less than $3$.
\end{Lemma}

Proof: Let $S=\{\sigma_x,\sigma_y,\sigma_z\}$. We have 
$\sum_{u\in S} uau^\dagger=-a$ for all $a\in\mathfrak{su}(2)$. Let $X,Y$ be
the bipartition of $G_0(H_d)$. By conjugating all qubits in $X$ with elements of
$S$ we obtain $-H_d$. 
\hfill $\Box$

If $H$ contains only $\sigma_z\otimes\sigma_z$ then the overhead is $1$. This is
optimal since we have $\mu\ge 1$.

\end{document}